\documentclass[a4paper,fleqn]{cas-sc}

\usepackage[authoryear]{natbib}

\def\tsc#1{\csdef{#1}{\textsc{\lowercase{#1}}\xspace}}
\tsc{WGM}
\tsc{QE}

\begin{document}
\let\WriteBookmarks\relax
\def\floatpagepagefraction{1}
\def\textpagefraction{.001}

\shorttitle{}    

\shortauthors{}  

\title [mode = title]{Airway Segmentation Network for Enhanced Tubular Feature Extraction}  



%

\author[1]{Qibiao WU}

\ead{241260091@st.usst.edu.cn}

\affiliation[a]{organization={School of Optical-electrical and Computer Engineering},
            addressline={University of shanghai for science and technology}, 
            city={Shanghai},
            postcode={200093}, 
            country={China}}

\author[a]{Yagang WANG}

\cormark[1]

\ead{ygwang@usst.edu.cn}

\author[b]{Qian ZHANG}

\ead{zhangqian@jsou.edu.cn}

\affiliation[b]{organization={School of Information Technology},
    addressline={Jiangsu Open University},
    city={Nanjing},
    postcode={210036},
    country={China}}




\begin{abstract}
Manual annotation of airway regions in computed tomography images is a time-consuming and expertise-dependent task. Automatic airway segmentation is therefore a prerequisite for enabling rapid bronchoscopic navigation and the clinical deployment of bronchoscopic robotic systems. Although convolutional neural network methods have gained considerable attention in airway segmentation, the unique tree-like structure of airways poses challenges for conventional and deformable convolutions, which often fail to focus on fine airway structures, leading to missed segments and discontinuities. To address this issue, this study proposes a novel tubular feature extraction network, named TfeNet. TfeNet introduces a novel direction-aware convolution operation that first applies spatial rotation transformations to adjust the sampling positions of linear convolution kernels. The deformed kernels are then represented as line segments or polylines in 3D space. Furthermore, a tubular feature fusion module (TFFM) is designed based on asymmetric convolution and residual connection strategies, enhancing the network’s focus on subtle airway structures. Extensive experiments conducted on one public dataset and two datasets used in airway segmentation challenges demonstrate that the proposed TfeNet achieves more accuracy and continuous airway structure predictions compared with existing methods. In particular, TfeNet achieves the highest overall score of 94.95\% on the current largest airway segmentation dataset, Airway Tree Modeling(ATM22), and demonstrates advanced performance on the lung fibrosis dataset(AIIB23). The code is available at https://github.com/QibiaoWu/TfeNet. 
\end{abstract}




\maketitle

\section{Introduction}
Pulmonary diseases represent a significant global health burden, accounting for tens of thousands of deaths annually. As the primary organ responsible for respiration, the lung is particularly susceptible to airborne pathogens (e.g., viruses, bacteria) and toxic pollutants. These exposures can lead to life-threatening conditions, including lung cancer, chronic obstructive pulmonary disease (COPD), and asthma. This underscores the imperative for advanced airway health monitoring. Currently, slice-wise annotated airways extracted from computed tomography (CT) scans serve as foundational components for preoperative pulmonary pathology diagnosis \citep{Cho2024UseOT}, bronchoscopic navigation systems \citep{Sganga2018OffsetNetDL,Shen2015RobustCL,Fried2023LandmarkBB} and   robotic bronchoscopy control systems \citep{Zou2022RoboticAssistedAO, Zhang2024ANH}  (Figure 1(a)). However, manual segmentation of airway remains a time-consuming and experience-dependent process, with reported durations exceeding 7 hours per patient CT \citep{Tschirren2009AirwaySF}. Such latency poses a significant barrier to rapid therapeutic interventions and impedes the deployment of modern bronchoscopic technologies. Over the past two decades, airway extraction methodologies have evolved through three distinct phases, including traditional techniques (e.g., thresholding, morphological operations) \citep{Lo2012ExtractionOA}, early machine learning approaches, and contemporary deep learning paradigms. Early research prioritized deep airway penetration while minimizing large-scale segmentation leakage, establishing critical benchmarks for subsequent algorithmic development.
Owing to the robust generalization capabilities and big-data processing advantages of deep learning, tracheal segmentation is relatively straightforward. The core challenge now centers on balancing airway continuity and segmentation accuracy. However, the core challenge in airway segmentation remains to improve segmentation accuracy while preserving structural continuity. This task is primarily constrained by three key issues: 1) inter-class imbalance between foreground (airway) and background voxels in CT images, 2) intra-class imbalance in voxel distribution between large and small airways within the foreground, and 3) gradient instability during model training. First, due to the overwhelming dominance of background voxels over airway voxels, models tend to be biased toward predicting non-airway regions, thereby compromising accurate airway segmentation. Second, within the airway structure itself, the disparity in voxel counts between major bronchi and smaller distal branches leads to underrepresentation of fine structures during training, increasing the likelihood of missed detections. Finally, gradient vanishing or exploding frequently occurs during deep learning model training, which disrupts stable parameter updates and results in a significant number of false negatives and false positives in the final segmentation outputs. As illustrated in Figure 1(c), these issues commonly induce missed detection and fragmentation of distal airways, compromising the topological integrity of the airway tree. Crucially, in pulmonary navigation and robotic systems, distal airway accessibility serves as a key metric for evaluating system advancement and clinical efficacy.

\begin{figure}
    \centering
    \includegraphics[width=0.5\linewidth]{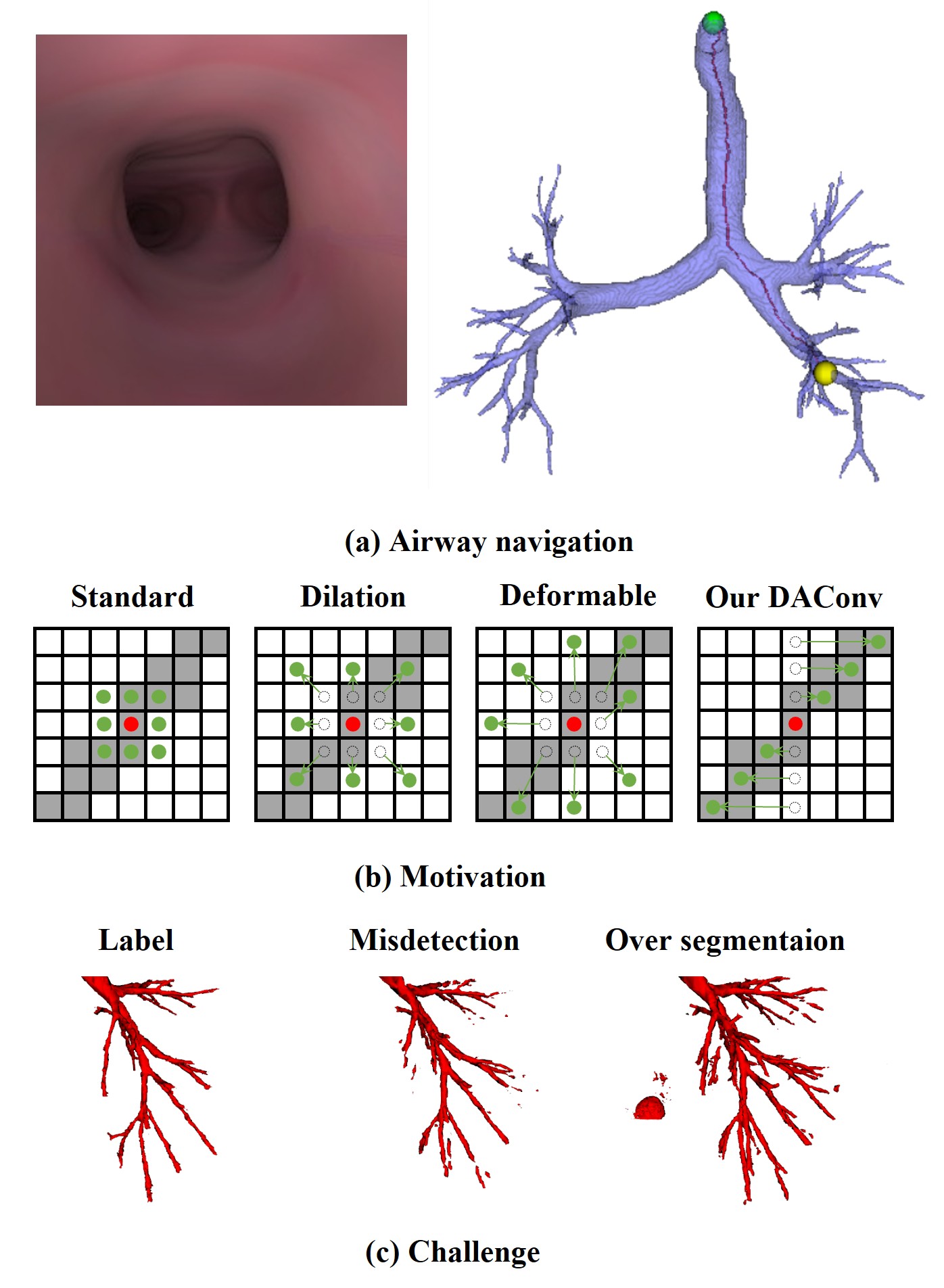}
    \caption{Application, Motivation and Challenges. (a) Airway segmentation can provide the rendering basis for airway virtual navigation and positioning. (b) Comparison of sampling locations for standard, dilated, and deformable convolutions with direction-aware convolutions. Our method is more capable of focusing on slender structures. (c) Airway segmentation is faced with misdetection and over-segmentation of the airway.}
    \label{fig:enter-label}
\end{figure}

To address the issue of class imbalance, some studies have proposed a multi-stage airway segmentation strategy \citep{Wu2022TwostageCT, Zhang2025DualphaseAS}, and dividing the binary airway labels into multiple categories \citep{Wang2023AccurateAT}, which allows the model to focus on different levels of airways. In terms of the loss function, Weight based on airway distance prior \citep{Zheng2020AlleviatingCG} and local-sensitive distance loss are integrated to balance the gradients \citep{Zhang2023TowardsCP}. Additionally, iterative training strategies have been applied to resolve issues related to airway discontinuities and missing segments, such as the fractured attention strategy based on airway discontinuity \citep{Wang2023AccurateAT} and the knowledge distillation method using a teacher-student network for unlabeled airways \citep{Wang2022NaviAirwayAB}, but this training method is extremely time-consuming. Some studies have also employed channel-wise fuzzy attention layers to alleviate intra-class heterogeneity and inter-class homogeneity \citep{Nan2022FuzzyAN}. Furthermore, research has shown that expanding the receptive field aids in the extraction of airway features \citep{Wu2022TwostageCT, Cheng2021SegmentationOT}.

This work conducts research from the perspective of airway feature extraction. As illustrated in Figure 1(b), the sampling positions of conventional convolution kernels are fixed, which does not align with the tree-like structure of the airways. While dilated convolutions expand the receptive field, they still fail to focus on extracting features from tubular structures. In contrast, deformable convolutions adaptively optimize sampling positions through learned offsets, enabling kernels shape to conform to target anatomies. Moreover, for segmentation tasks involving fine tubular structures (e.g., vascular, coronary, airway trees), it is crucial to maintain the continuity of the convolution kernels and avoid divergence in the convolution sampling positions. Inspired by deformable convolution, we propose Direction-Aware Convolution (DAConv), which is a novel operator that dynamically rotates sampling grids according to learnable angles. This generates polyline-optimized sampling patterns that significantly enhance feature capture capabilities for tree-like structures. Further, we introduce a plug-and-play Tubular Feature Fusion Module (TFFM) which combines DAConv kernels along the x, y, and z axes, employs residual pathways to fuse local directional features with global contextual priors, Enhances the intensity of local airway feature. Based on TFFM, we propose a novel tubular feature extraction network, named TfeNet, for airway segmentation. 
The main contributions of this work are as follows:

1.	We analyze the common issues present in airway segmentation methods and propose a novel tubular feature extraction network named TfeNet. It address these issues from the prespective of airway feature extraction, effectively improving the segmentation capability.

2.	We introduce Direction-Aware Convolution (DAConv) to replace standard and deformable convolutions, enhancing feature extraction for tree-like structures via learnable angel parameters.

3.	We propose a plug-and-play Tubular Feature Fusion Module (TFFM) leveraging multi-axis DAConv layers. This module first aggregates local direction-aware features, then employs residual connections to preserve original features and stabilize training dynamics.

4.	TfeNet is evaluated on multiple benchmark datasets to demonstrate its effectiveness. Notably, on the largest airway segmentation dataset, Airway Tree Modeling, the proposed method achieves the highest average score when compared with state-of-the-art approaches. Furthermore, it demonstrates superior performance on the BAS, ATM22 and AIIB23.




\section{Related work}
\subsection{Airway segmentation}
Convolutional Neural Network (CNN)-based architecture has been widely adopted in the field of medical image segmentation. profits from its innovative encoder-decoder structure and skip-connection mechanism, 3D U-Net has gained extensive popularity \citep{iek20163DUL}. Meanwhile, nnUNet \citep{Isensee2020nnUNetAS} which forgoes complex architectural innovations, focuses instead on systematic optimization of data preprocessing and training strategies, making it a popular baseline method among researchers in the medical imaging community. For the task of airway segmentation, Qin et al. reformulate the problem as predicting connectivity across 26 neighboring voxels, estimating the connectivity between each voxel and its 26 surrounding neighbors \citep{Qin2019AirwayNetAV, Qin2020AirwayNetSEAS}. They further introduce mechanisms incorporating voxel coordinates, lung boundary distance maps, and multi-scale contextual fusion. Additionally, Qin et al. propose a feature recalibration module and an attention distillation module, which significantly enhances the network's ability to learn features specific to tubular airway structures \citep{Qin2020LearningTC}. Zhang et al. develop a collaborative feature decoupling framework that synergizes features from clean and noisy domains via a bias-aware discriminative encoder, enhancing feature robustness through semi-supervised learning \citep{Zhang2022CFDACF}. Zheng et al. introduce the WingsNet architecture, which employs group-wise supervision to provide complementary gradient flows to both the encoder and decoder, thereby mitigating the issues of gradient erosion and gradient explosion during network training \citep{Zheng2020AlleviatingCG}. Wu et al. incorporate a 3D contextual Transformer module into their model, enabling effective capture of contextual and long-range dependencies in CT images \citep{Wu2022TwostageCT}. However, none of these networks have explicitly explored the extraction of airway continuity features, which remains a critical challenge in airway segmentation.
\subsection{Deformable convolution}
Airway segmentation is inherently challenging due to the extreme class imbalance between background and foreground regions, as well as between large and small airways \citep{Zheng2020AlleviatingCG}. Deformable convolution addresses this challenge by enabling dynamic deformation of the sampling locations within the convolution kernel, thereby adapting to diverse feature shapes and enhancing feature extraction capabilities \citep{Dai2017DeformableCN}. Zhu et al. introduce a learnable modulation term into deformable convolution, allowing for weighted contributions from each sampling location to the convolution kernel \citep{Zhu2018DeformableCV}. Jin et al. integrate deformable convolution into the U-Net architecture, significantly improving the segmentation performance of thin tubular structures \citep{Jin2018DUNetAD}. Yu et al. propose the concept of holistic deformable convolution, which learns offsets for the entire convolution kernel to prevent feature dispersion during the extraction process \citep{Yu2022EntireDC}. Feng et al. further develope an adaptive deformable convolution by introducing learnable dilation rate parameters that jointly model the modulation and offset learning processes \citep{Chen2020AdaptiveDC}. Reza et al. combine deformable convolution with a large-kernel attention mechanism, leading to enhanced feature representation \citep{Azad2023BeyondSD}. However, the subtle characteristics of small airways are often lost after only a few convolutional layers, resulting in incomplete reconstruction and prediction of these regions during later stages of the network \citep{Zheng2021RefinedLW}. While deformable convolution offers flexibility, its learned offsets may not adequately focus on fine-grained features. To better capture tubular and fine structures, Qi et al. propose dynamic snake convolution(DSConv), which imposes continuity constraints on offset learning and fuses features across different receptive field perspectives within the convolution kernel \citep{Qi2023DynamicSC}. Tian et al. further extend DSConv by introducing a Coordinate Attention module based on snake convolution, which improves the network's ability to extract fine-scale features \citep{Yao2024ObjectRN}. Despite these advances, none of the aforementioned methods explicitly model the tree-like hierarchical structure inherent to airway anatomy.

\section{Methods}
In this study, we propose DAConv to capture tubular features in spatial domains, as well as the Tubular Feature Fusion Module (TFFM), which integrates local tubular feature. The TFFM is embedded within an encoder-decoder network architecture, forming the airway segmentation network termed TfeNet, which is specifically designed to enhance the extraction of tubular features. The overall architecture of the network is illustrated in Figure 2(a), where skip connections are established between the Encoder and Decoder to preserve spatial details and facilitate gradient flow. $x\in X$ is the input volume and $y\in Y$ is the ground-truth of segmentation mask. We design a model $\hat{y}=\operatorname{TfeNet}(x, \boldsymbol{\theta})$, where $\theta$ represents the model parameters and $\hat{y}$ is the likelihood map of prediction. The proposed TFFM is illustrated in Figure 2(b). Given that employing TFFM in the final layer of the network would lead to an increase in the number of parameters and a decrease in computational speed without enhancing performance, the TFFM is replaced with a ResConv module (as shown in Figure 2(c)). This substitution maintains the network's efficiency without compromising its effectiveness. After each convolution, Instance Normalization is applied for normalization, followed by the ReLU activation function to introduce non-linearity. For clarity, these processing steps are not depicted in Figure 2.

\begin{figure}
    \centering
    \includegraphics[width=1\linewidth]{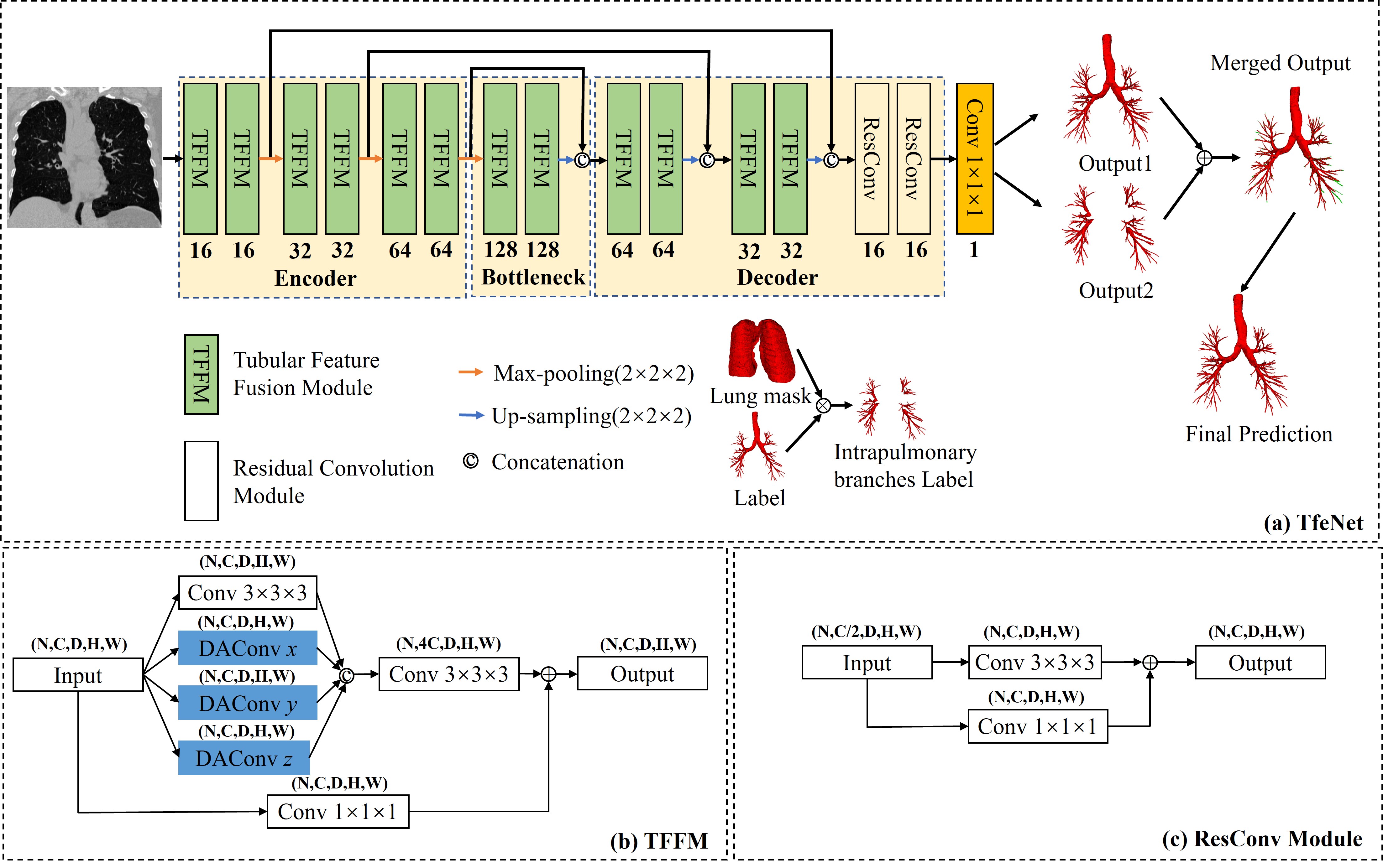}
    \caption{This figure illustrates the detailed architecture of the proposed TfeNet.}
    \label{fig:enter-label}
\end{figure}
\subsection{DAConv: Direction-aware convolution}
Traditional convolutional kernels and general deformable convolutions struggle to effectively capture the tree-like structural features of airways. For deformable convolutions, there is a notable limitation in capturing fine structures; they can even lead to divergent sampling locations, causing the network to fail to converge. Inspired by tree structures, it is proposed to use combinations of "I"-shaped and "V"-shaped configurations in space to represent branches, resembling a "Y" shape. This approach allows for a more accurate depiction of the intricate and branching nature of airway structures, addressing the limitations of conventional and deformable convolutions in capturing these detailed morphological features. To sum up, we propose DAConv for capturing airway dendritic features. DAConv is defined as three different variants of linear convolution, namely the variants of convolution kernels $k \times 1 \times 1$, $1 \times k \times 1$, and $1 \times 1 \times k$ in the x, y, and z axes,where $k\in \left \{ 1,3,5,7,\dots  \right \}$. The deformation process of DAConv is illustrated along the x-direction as an example. Given the linear convolution kernel sampling locations along the x-direction, denoted as:
\begin{equation}
K = \left\{ (x-1, 1, 1), (x-2, 1, 1), \cdots, (x+1, 1, 1), (x+2, 1, 1) \right\}
\end{equation}

To enhance the flexibility of the linear convolution and better capture the complex structural features of tree-like objects in space while maintaining kernel continuity, we introduce a rotation parameters $\theta$.
\begin{equation}
\theta = \begin{bmatrix} \theta_1 & \theta_2 & \theta_3 & \theta_4 \end{bmatrix}
\end{equation}

The rotation matrices of a point in space around the x-axis, y-axis, and z-axis are, respectively:
\begin{equation}
\left\{
\begin{aligned}
R_x(\theta) &= \begin{bmatrix}
1 & 0 & 0 \\
0 & \cos\theta & -\sin\theta \\
0 & \sin\theta & \cos\theta
\end{bmatrix} \\
R_y(\theta) &= \begin{bmatrix}
\cos\theta & 0 & \sin\theta \\
0 & 1 & 0 \\
-\sin\theta & 0 & \cos\theta
\end{bmatrix}, \\
R_z(\theta) &= \begin{bmatrix}
\cos\theta & -\sin\theta & 0 \\
\sin\theta & \cos\theta & 0 \\
0 & 0 & 1
\end{bmatrix}
\end{aligned}
\right.
\end{equation}

We denote the specific sampling locations of each grid in the x-direction $k \times 1 \times 1$ linear convolution as $K_{i \pm c} = (x_{i \pm c}, y_{i \pm c}, z_{i \pm c})$ where $c=0,1,2,3,...$ denotes the distance between the initial sampling location and the center grid of rotation, before rotation $y_{i \pm c} = z_{i \pm c} = 1$ The convolution kernel is divided into two segments with the rotation center located at $(k-1)/2$ , where $k$ denotes the length of the linear convolution kernel. Each segment is first rotated around the y-axis and then around the z-axis by a certain angle, the sampling location offset can be obtained through the rotation matrix as:
\begin{equation}
\Delta P_x = c R_y(\theta_a) R_z(\theta_b)
\end{equation}
\begin{equation}
\Delta P_x = 
\left\{
\begin{aligned}
(\Delta x_{i-c}, \Delta y_{i-c}, \Delta z_{i-c}) &= (-c \cos(\theta_1) \cos(\theta_2), -c \cos(\theta_1) \sin(\theta_2), c \sin(\theta_1)) \\
(\Delta x_{i+c}, \Delta y_{i+c}, \Delta z_{i+c}) &= (c \cos(\theta_3) \cos(\theta_4), c \cos(\theta_3) \sin(\theta_4), -c \sin(\theta_3))
\end{aligned}
\right.
\end{equation}
Therefore, the final sampling locations can be expressed as:
\begin{equation}
K_{i \pm c} = 
\left\{
\begin{aligned}
(x_{i-c}, y_{i-c}, z_{i-c}) &= \left( x_i + d_x (k-1)/2 + \Delta x_{i-c}, y_i + \Delta y_{i-c}, z_i + \Delta z_{i-c} \right) \\
(x_{i+c}, y_{i+c}, z_{i+c}) &= \left( x_i + d_x (k-1)/2 + \Delta x_{i+c}, y_i + \Delta y_{i+c}, z_i + \Delta z_{i+c} \right)
\end{aligned}
\right.
\end{equation}
Here $d_x$ is dilation of x-axis, it is set to 1 in this paper.
Since the newly computed sampling locations are typically fractional values, while image coordinates are usually integers, trilinear interpolation is employed to obtain the intensity values at these sampling points. The same operation is applied to the linear convolutions along the y-direction (rotated first around the x-axis and then around the z-axis) and the z-direction (rotated first around the x-axis and then around the y-axis), resulting in the updated sampling positions for each case.
\begin{equation}
K = \sum_{K'} C(K', K) \cdot K'  
\end{equation}
Here $K$ are a fractional coordinate, $K'$ enumerates all integral spatial locations and $C$ is the trilinear interpolation kernel.

As illustrated in Figure 3(a), the learning process of the rotation angle $\theta$ is as follows: first, a $3\times3\times3$ convolutional layer is applied, followed by instance normalization. The output is then passed through a tanh activation function to scale the values to the range $[-1,1]$. Finally, a scaling factor $q$ is multiplied to obtain the rotation angle within the range $[-q, q]$, where $q \in 
(0, \pi]$. In this study, $q$ is set to a specific value $\pi/4$.
\begin{equation}
Q = q\cdot\tanh\left(\text{IN}\left(\text{Conv}_{3\times3\times3}(x)\right)\right)
\end{equation}
Here IN is instance normalization. As shown in Figure 3(c), the radius of rotation now covers the entire spherical region, with low overlap of edge features. Taking a $k = 7$ linear convolution in the x-direction as an example, the schematic diagram after the convolution kernel undergoes rotational transformation is depicted in Figure 3(b). Initially, the sampling positions of the convolution kernel are indicated in blue, and the center of rotation is marked by a yellow point. After the first round of rotation, these positions transform to those indicated in orange, and following a second round of rotation, they change to those shown in black.  It is visually apparent that the flexibility of the rotated convolution kernel has been enhanced.Notably, when $\theta_1=\theta_3$ and $\theta_2=\theta_4$, the shape of the convolution kernel represents a line segment in space.
\begin{figure}
    \centering
    \includegraphics[width=0.5\linewidth]{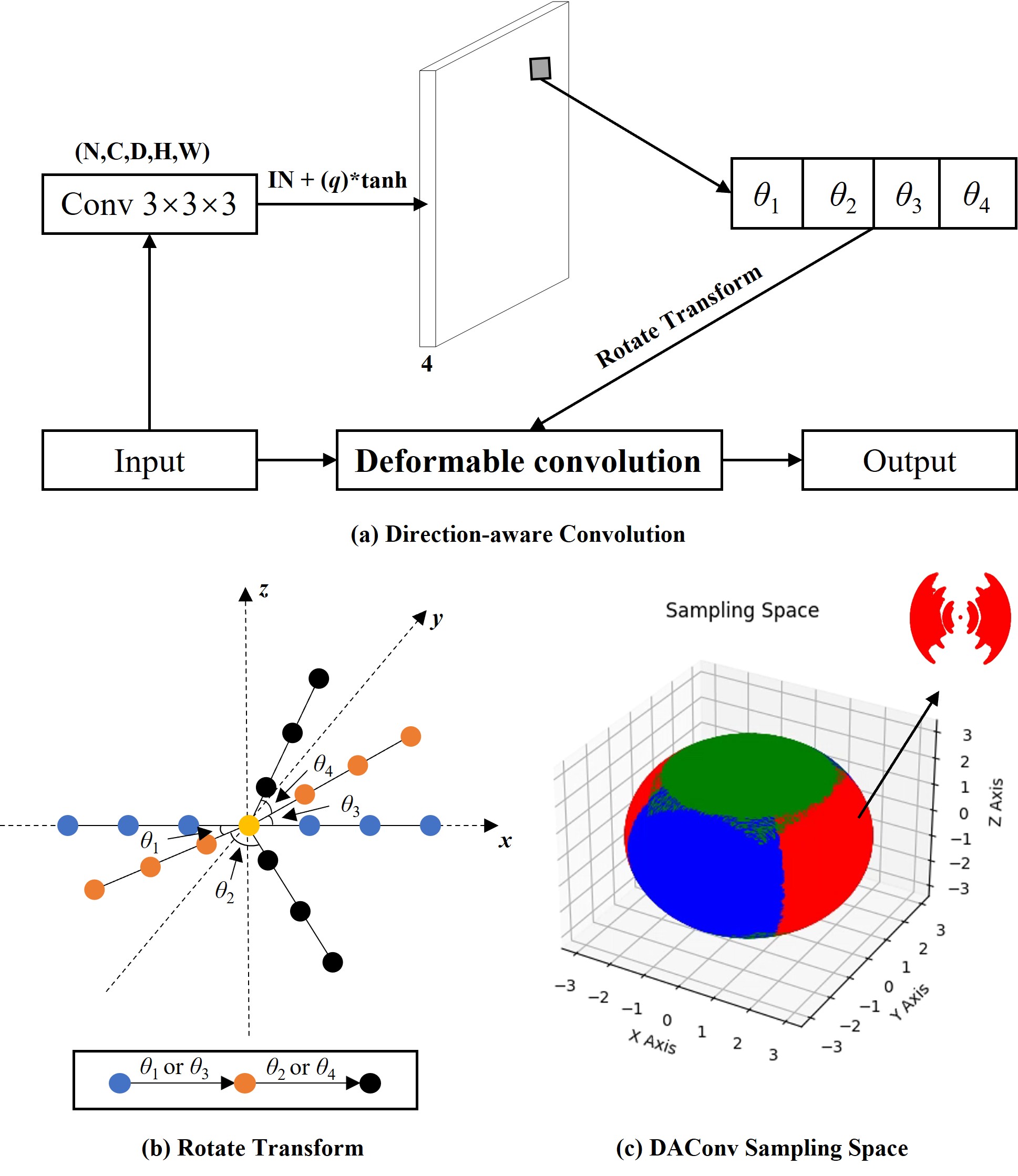}
    \caption{The detailed flowchart of the DAConv process are as follows: (a) The rotation process of DAConv convolution; (b) Initially, the sampling positions of the convolution kernels are represented by blue, with the rotation center marked by a yellow point. After the first rotation, the positions become orange, and after the second rotation, they turn black; (c) The sampling space of DAConv, where the red, blue, and green lines represent the deformable coverage regions of convolution in the x, y, and z directions, respectively. These three directions cover a spherical receptive field, with a low degree of overlap at the edges.}
    \label{fig:enter-label}
\end{figure}

\subsection{TFFM: Tubular feature fusion module}
Inspired by asymmetric convolutions \citep{Ding2019ACNetST} and residual connections \citep{He2015DeepRL}, we propose a tubular feature extraction module based on the designed DAConv. This module integrates the local feature maps from DAConv in the x, y, and z directions with those from a $3\times3\times3 $ convolution to capture the local contextual information of the input feature map:
\begin{equation}
f = \text{cat}\left(\text{DAConv}_x(x), \text{DAConv}_y(x), \text{DAConv}_z(x), \text{Conv}_{3\times3\times3}(x)\right)
\end{equation}
Here cat refers to the concatenation operation along the channel dimension, resulting in the fused feature $f$. To ensure that input information is preserved, a residual connection is introduced, and the feature map is generated through element-wise addition.
\begin{equation}
y_{tf} = \text{Conv}_{3\times3\times3}(f) \oplus \text{Conv}_{1\times1\times1}(x)
\end{equation}

The TFFM is integrated in an Encoder-Decoder network similar to 3D-UNet. It is a plug-and-play module, allowing for easy replacement of convolutional layers within the other networks.

\subsection{Loss function}
The General Union Loss (GUL) \citep{Zheng2020AlleviatingCG} has been proven to effectively alleviate the intra-class and inter-class imbalance in airways. Therefore, this study uses this loss for training to address the intra-class and inter-class imbalance problem in airway segmentation. The expression for GUL is as follows:
\begin{equation}
U = 1 - \frac{\sum_{i=1}^{N} \omega_i p_i^{l_i} g_i}{\sum_{i=1}^{N} \omega_i (\alpha p_i + \beta g_i)}
\end{equation}
Here $\omega$ are weight, $\alpha+\beta=1$, $r_l=0.7$ are root of prediction.

The size of airway branches can be quantified through local class imbalance between foreground and background voxels, as measured by the local foreground ratio metric:
\begin{equation}
FR_p = \frac{1}{N} \sum_{x_i \in B} y_i
\end{equation}

We generally desire that small airway regions have a larger weight; thus, a Local-imbalance-based Weight(LIB Weight) \citep{Zheng2021RefinedLW} can be designed:
\begin{equation}
\omega_p = (1 - \lambda) \left( f_c \left( -\log_{10} FR_p \right) \right)^r + \lambda
\end{equation}
Here $f_c=\text{MIN}(x,1)$ is applied to prevent the exponential growth of the values greater than 1. We chooses $\lambda=0.05$, and the root r randomly from 2 to 3.

During the validation process of the ATM22 dataset, the weight assigned by the LIB Weight to the large airways was set too low, and TfeNet became more biased towards training fine-grained segmentation, leading to partial missed detections at the starting position of the trachea. Therefore, Tversky loss \citep{Salehi2017TverskyLF} was employed for training in the Output1 phase of this dataset:
\begin{equation}
T = 1 - \frac{\sum_{i=1}^{N} p_i g_i}{\sum_{i=1}^{N} (\alpha p_i + \beta g_i)}
\end{equation}
\section{Dataset and Implementation}
\subsection{Datasets}
In our work, three datasets are used for evaluation, encompassing one public dataset and two airway segmentation challenges, As shown in Table 1.
\label{}
\begin{table}
\caption{Airway segmentation dataset details}\label{tbl1}
\begin{tabular}{l c c c c c c}
\toprule
Datasets & Total & Training & Validation & Test & Slice thickness ranges & Spatial resolution ranges \\
\midrule
BAS \citep{Zheng2020AlleviatingCG} & 90 & 50 & 20 & 20 & 0.50mm to 1.80mm & 0.51mm to 0.82mm \\
ATM22 \citep{Zhang2023MultisiteMA} & 500 & 300 & 50 & 150 & 0.51mm to 1.00mm & 0.51mm to 0.92 mm \\
AIIB23 \citep{Nan2023HuntingIB} & 285 & 120 & 52 & 113 & 0.40mm to 2.00mm & 0.42mm to 0.93mm \\
\bottomrule
\end{tabular}
\end{table}

\textbf{Binary Airway Segmentation Dataset (BAS)}: The BAS dataset\citep{Zheng2020AlleviatingCG} consists of 90 CT scans with airway annotations, including 70 cases from LIDC \citep{Armato2011TheLI} and 20 cases from the EXACT'09 \citep{Lo2012ExtractionOA} training set. The spatial resolution ranges from 0.51mm to 0.82mm, with slice thickness varying from 0.50mm to 1.80mm. For the BAS dataset, We divided the data set in the same way as in reference \citep{Zheng2020AlleviatingCG}, dividing 90 CT scans into training set (50 scans), validation set (20 scans) and test set (more than 20 scans). This dataset has been reported in Wang's study  \citep{Wang2023AccurateAT} to have incomplete labels, such as inaccurate small airway annotations and blurred airway boundary markings, which may cause a conflict between high topological metrics and high accuracy metrics.

\textbf{Airway Tree Modeling Dataset (ATM22):} The ATM’22 challenge \citep{Zhang2023MultisiteMA} consists of a large-scale dataset with 500 chest CT scans (300 for training, 50 for validation, and 150 for testing). There was one incorrectly labeled sample in the training set, which is removed by the official team. Due to the crop size of 128, one sample with insufficient slices (fewer than 128) is manually removed, resulting in a final training set size of 298. The data sources include EXCAT’09 \citep{Lo2012ExtractionOA}, LIDC \citep{Armato2011TheLI}, and Shanghai Chest Hospital, covering both healthy subjects and patients with severe lung diseases. The in-plane spatial resolution of the CT scans ranges from 0.51 to 0.92 mm, and the slice thickness is between 0.51 and 1.00 mm. We train our model on the training set and submit our results for evaluation on the test set. Then, the predictions from three models were combined using majority voting to obtain the initial segmentation results. These initial labels were manually checked and edited by three radiologists using the same annotation principles.

\textbf{Airway-Informed Quantitative CT Imaging Biomarker for Fibrotic Lung Disease Dataset (AIIB23):} The AIIB23 challenge \citep{Nan2023HuntingIB,Li2022HumanTT} is the first airway segmentation challenge focused on pulmonary fibrosis. The CT scans for this dataset were sourced from the Royal Brompton Hospital and Imperial College London. The AIIB23 dataset consists of 285 cases, with 120 for training, 52 for validation, and 113 for testing. Among these, 235 cases are from patients with fibrotic lung diseases, and 50 cases are from COVID-19 patients. It is worth noting that AIIB23 has 512×512 and 768×768 slice resolution. The spatial resolution ranges from 0.42mm to 0.93mm, with slice thickness varying from 0.40mm to 2.00mm. The airway structures in these scans are meticulously annotated by three experienced radiologists. In images from patients with fibrotic lung diseases, the distal airways exhibit dilation, and honeycomb-like cavities are observed. These cavities have voxel intensities similar to those of the airways, which increases the difficulty of airway segmentation.

\subsection{Implementation details}
In the data preprocessing stage, a pre-trained U-Net (R-231) model is first employed to obtain the lung mask \citep{Hofmanninger2020AutomaticLS}. Based on this lung mask, the internal airway labels are extracted, and the bounding box of the lung mask is used to define the target volume. The Hounsfield Unit (HU) values of the region of interest are clipped to the range [–1000, 600] and then normalized to $[0, 1]$ after data augmentation. The training process follows a two-stage strategy[9], referred to as Stage Output1 and Stage Output2. Due to GPU memory constraints, patch-based training is adopted. In each epoch, 16 patches of size $128\times128\times128$ are randomly sampled from each patient, with a batch size of 1. In DAConv, the parameters are set as $q=\pi/4$, $k=7$. Stochastic Gradient Descent (SGD) is used as the optimizer, with a momentum coefficient of 0.9 and an initial learning rate of 0.01. Both Output1 and Output2 stages are trained for 60 epochs. The learning rate is multiplied by 0.1 at the 20th and 40th epochs. To prevent large angular updates that may cause network instability, the learning rate for DAConv layers in the Encoder and Bottleneck is reduced by a factor of 0.1 compared to the base learning rate. The Decoder maintains the same learning rate as the base to better recover airway features. The weights and biases of the angle learning layers in DAConv are initialized to 0, and the segmentation threshold is set to 0.5. Data augmentation is performed using random rotation. In Stage 1, the rotation augmentation threshold is set to 0.7, with GUL parameter$\alpha=0.05$. Stage 2, the rotation threshold is increased to 0.9, and $\alpha$ is adjusted to 0.1. During validation and testing, a sliding-window strategy is applied to extract patches of size $128\times128\times128$ with a stride of $64\times64\times64$. Post-processing includes extracting the largest connected airway component and filling any holes. In the ATM22 challenge, the Output1 stage is trained using the Tversky Loss to reduce false negatives in large airways. nnUNet demonstrates strong performance in spatial resampling and dimension normalization; therefore, it is selected for training the Output1 stage to address variations in slice thickness and spatial resolution across the AIIB23 dataset. The Output2 stage of TfeNet is trained only once, with $\alpha=0.1$ and a rotation augmentation threshold of 0.9. All experiments are conducted on an RTX 3090 GPU with CUDA 12.1, CuDNN 12.1, and PyTorch 2.4.1.
\subsection{Evaluation metric}
In this study, we will use Precision, Dice Similarity Coefficient (DSC), Tree Length Detected Rate (TD), and Branch Detected Rate (BD) to evaluate airway segmentation. Precision and DSC will measure the accuracy of the segmentation, while TD and BD will assess the topological correctness of the segmentation. Among these. We denote the true positive volume part as TP and true negative volume part as TN. Similarly, FN denotes the false negative volume part and FP is the false positive volume part. \textit{Precision} reflects the accuracy of the segmentation:
\begin{equation}
Precision = \frac{|TP|}{|TP + FP|}
\end{equation}
Here the $|\cdot|$ denotes the sum operation that returns the number of voxels.

\textit{DSC} reflects the degree of similarity between the segmentation result and the ground-truth labels:
\begin{equation}
DSC = \frac{|2 \times TP|}{|FP + 2 \times TP + FN|}
\end{equation}

\textit{TD} is defined as the ratio of the detected tree length in the segmentation to the actual tree length of the airway in the ground-truth:
\begin{equation}
TD = T_{det}/T_{ref}
\end{equation}
Here $T_{det}$ denotes the total length of all branches detected in the prediction, and $T_{ref}$ represents the whole tree length in the ground-truth.

\textit{BD} represents the percentage of correctly detected airway branches compared to the total number of branches in the ground-truth:
\begin{equation}
BD = B_{det}/B_{ref}
\end{equation}
Here $B_{det}$ denotes the total correct branches detected in the prediction, and $B_{ref}$ represents the whole number of branches in the ground-truth. Consistent with the rules in ATM22 challenge, a branch in the prediction is identified as ‘correct’ only if more than 80\% of centerline voxels extracted from the certain branch are within the ground-truth. the accuracy and continuity metric of the airways are inversely proportional. To ensure fairness, in the ATM22 challenge, the airway accuracy and continuity metrics are evaluated using an average score, calculated as follows:
\begin{equation}
MeanScore=0.25\times(Precision + DSC + TD + BD)
\end{equation}

\textit{Leakage}, \textit{IOU} and \textit{OverallScore} were also used in the AIIB23 adjustment to evaluate airways. The Leakage is defined as the proportion of total false positive volumes with respect to the ground truth volumes:
\begin{equation}
\text{Leakage} = 1 - \frac{|FP|}{|TP + FN|}
\end{equation}

\textit{IOU} is defined as intersection over union score:
\begin{equation}
IOU = 1 - \frac{|TP|}{|TP + FP + FN|}
\end{equation}

\textit{OverallScore} is the overall score of the validation set of the AIIB23 challenge, which is expressed as follows:
\begin{equation}
OverallScore = (IOU + Precision + TD + BD) \times 0.25 \times 0.7 + Leakages \times 0.3
\end{equation}

\section{Experiments and results}
\subsection{Results on BAS}
To validate the superiority of the proposed network in extracting tree-like structural features and detecting airway topology, we conduct comparative experiments on the BAS dataset against state-of-the-art airway segmentation networks. The results are summarized in Table 2. All methods employed Dice Loss and rotation-based data augmentation with a threshold of 0.7. To ensure a fair comparison, all networks are configured with the same learning rate, and an input patch size of $128\times128\times128$ was used. nnUNet adopt its recommended preprocessing pipeline, and we adjusted the number of network layers from 6 to 4 for compatibility. DCUNet (Deformable Convolution UNet) is implemented as a control experiment using deformable convolutions, based on the 3D U-Net architecture, one of the convolution kernels in each double convolution block was replaced with a deformable convolution kernel to compare the performance differences between deformable convolution and DAConv. Due to the excessive GPU memory consumption of the original PyTorch implementation of DSConv on 3D datasets, we reimplement it using CUDA programming, and the corresponding code has been made publicly available in our repository. The kernel size is set to 7. The proposed TfeNet achieves excellent accuracy while attaining the highest MeanScore of 87.37\%. It also demonstrated the best performance in capturing airway topology, with TD and BD scores of 84.27\% and 77.51\%, respectively, representing improvements of 2.32\% and 3.61\% over nnUNet. Although nnUNet achieves the highest DSC coefficient of 93.87\% due to its robust pipeline design, it underperform in topological preservation compared to our method. DCUNet incorporates deformable convolutions but lacks continuity constraints, leading to divergent sampling locations and poor focus on airway regions, resulting in relatively low TD and BD scores of 66.68\% and 57.23\%. Nevertheless, it achieves the best Precision score of 95.95\%. The COT module in CotUNet extends the original convolution's capability to capture global receptive field information, enabling the network to learn richer contextual features. In contrast, DSCNet’s feature fusion module does not retain input information through residual connections. As the network depth increases, the original airway patterns gradually diminish, limiting its performance in preserving fine-scale structures.
\label{}
\begin{table}
\caption{Comparison of airway segmentation networks on BAS test set. The best results are in bold.}\label{tbl1}
\begin{tabular}{l c c c c c}
\toprule
Methods & TD(\%) & BD(\%) & Precision(\%) & DSC(\%) & MeanScore(\%) \\
\midrule
3D-UNet \citep{iek20163DUL} & $76.86 \pm 15.81$ & $67.67 \pm 15.34$ & $95.08 \pm 2.21$ & $92.65 \pm 3.16$ & $83.07 \pm 9.13$ \\
nnUNet \citep{Isensee2020nnUNetAS} & $81.95 \pm 10.65$ & $73.90 \pm 13.12$ & $95.77 \pm 1.95$ & $\mathbf{93.87 \pm 1.97}$ & $86.37 \pm 6.92$ \\
Qin \citep{Qin2020LearningTC} & $64.70 \pm 20.27$ & $54.20 \pm 19.17$ & $95.88 \pm 2.25$ & $89.29 \pm 7.19$ & $76.02 \pm 12.22$ \\
WingsNet \citep{Zheng2020AlleviatingCG} & $79.16 \pm 12.73$ & $70.78 \pm 13.58$ & $94.92 \pm 2.12$ & $92.70 \pm 2.24$ & $84.39 \pm 7.67$ \\
DCUNet \citep{Dai2017DeformableCN} & $66.68 \pm 23.68$ & $57.23 \pm 22.30$ & $\mathbf{95.95 \pm 2.37}$ & $85.57 \pm 17.02$ & $76.36 \pm 16.34$ \\
CotUnet \citep{Wu2022TwostageCT} & $80.18 \pm 13.29$ & $71.85 \pm 13.79$ & $94.59 \pm 2.44$ & $92.76 \pm 3.05$ & $84.85 \pm 8.14$ \\
DSCNet \citep{Qi2023DynamicSC} & $73.10 \pm 23.49$ & $63.98 \pm 23.45$ & $95.93 \pm 2.16$ & $90.52 \pm 7.23$ & $80.88 \pm 14.08$ \\
\midrule
TfeNet(Ours) & $\mathbf{84.27 \pm 12.97}$ & $\mathbf{77.51 \pm 14.20}$ & $94.54 \pm 2.38$ & $93.14 \pm 2.91$ & $\mathbf{87.37 \pm 8.12}$ \\
\bottomrule
\end{tabular}
\end{table}

The quantitative analysis of the proposed method compared to state-of-the-art airway segmentation methods is presented in Table 3. Our method achieves TD and BD scores of 95.48\% and 93.83\%, respectively, along with a Precision score of 90.79\%. Compared to the predictions from Output1, our Output2 configuration shows improvements of 1.99\% and 2.66\% for TD and BD, respectively, indicating that Output2 can learn more features of small airways. However, due to Output2 identifying additional unannotated airways, Precision decreased by 1.32\%. V-Net \citep{Milletar2016VNetFC}, Juarez \citep{Juarez2018AutomaticAS}, and AirwayNet \citep{Qin2019AirwayNetAV} demonstrate high Precision metrics of 97.81\%, 96.40\%, and 95.76\%, respectively, but exhibit relatively lower performance in topological metrics TD and BD. Conversely, CotUNet \citep{Wu2022TwostageCT}, and Zhang \citep{Zhang2023TowardsCP} achieves superior TD and BD scores of 94.90\% and 92.40\%, and 96.52\% and 91.50\%, respectively. However, CotUNet suffered from low Precision scores, manifesting as airway walls dilation, this phenomenon attributed to poor dataset annotation quality in Wang's study \citep{Wang2023AccurateAT}. Zhang manage to achieve high TD and BD scores while maintaining a respectable Precision of 91.24\%, thanks to the designed LSD and CAS modules, CAS module ensure voxel-wise prediction accuracy and repair disconnected airways through the LSD module. It is evident that achieving high topological metrics (TD and BD) and high accuracy simultaneously presents a challenge in the BAS dataset, often resulting in a trade-off between high topology with lower precision or vice versa. In contrast, our approach achieves a balanced performance across all these metrics.

\label{}
\begin{table}
\caption{Quantitative comparison of state-of-the-art airway segmentation methods on BAS test set. The best and next best results are in bold.}\label{tbl1}
\begin{tabular}{l c c c}
\toprule
Methods & TD(\%) & BD(\%) & Precision(\%) \\
\midrule
nnUNet \citep{Isensee2020nnUNetAS} & $86.84 \pm 7.00$ & $79.21 \pm 9.43$ & $94.36 \pm 2.34$ \\
V-Net \citep{Milletar2016VNetFC} & $33.96 \pm 17.96$ & $22.04 \pm 14.22$ & $\mathbf{97.81 \pm 1.50}$ \\
Wang \citep{Wang2019TubularSS} & $86.60 \pm 8.50$ & $83.50 \pm 11.20$ & $93.40 \pm 2.10$ \\
Juarez \citep{Juarez2018AutomaticAS} & $84.10 \pm 8.6$ & $82.10 \pm 12.40$ & $\mathbf{96.40 \pm 1.80}$ \\
Jin \citep{Jin20173DCN} & $85.40 \pm 10.40$ & $83.10 \pm 11.50$ & $93.90 \pm 1.90$ \\
AirwayNet \citep{Qin2019AirwayNetAV} & $83.59 \pm 10.37$ & $81.37 \pm 13.84$ & $95.76 \pm 1.84$ \\
Qin \citep{Qin2020LearningTC} & $91.82 \pm 5.31$ & $87.64 \pm 9.18$ & $91.54 \pm 2.89$ \\
BronchiNet \citep{Juarez2021AutomaticAS} & $64.28 \pm 16.28$ & $57.52 \pm 18.01$ & $85.88 \pm 6.21$ \\
Zhang \citep{Zhang2023TowardsCP} & $\mathbf{96.52 \pm 3.95}$ & $91.50 \pm 2.99$ & $91.24 \pm 2.78$ \\
WingsNet \citep{Zheng2020AlleviatingCG} & $92.50 \pm 4.50$ & $88.70 \pm 7.90$ & $91.40 \pm 3.30$ \\
CotUnet \citep{Wu2022TwostageCT} & $94.90 \pm 3.40$ & $\mathbf{92.40 \pm 6.00}$ & $86.90 \pm 4.10$ \\
Nan \citep{Nan2022FuzzyAN} & $92.71 \pm 7.93$ & $89.01 \pm 10.3$ & $91.87 \pm 1.50$ \\
NaviAirway \citep{Wang2022NaviAirwayAB} & $87.34 \pm 7.15$ & $80.99 \pm 9.50$ & $86.72 \pm 4.06$ \\
\midrule
TfeNet1(Ours) & $93.49 \pm 5.44$ & $91.17 \pm 7.28$ & $92.11 \pm 3.02$ \\
TfeNet2(Ours) & $\mathbf{95.48 \pm 3.53}$ & $\mathbf{93.83 \pm 5.20}$ & $90.79 \pm 3.27$ \\
\bottomrule
\end{tabular}
\begin{flushleft} 
     \textbf{Note:} TfeNet1 and TfeNet2 represent the results of testing with Output1, and Output1 combined with Output2, respectively.
\end{flushleft}
\end{table}

The case comparisons of state-of-the-art airway segmentation methods on the BAS test set are shown in Figure 4. The proposed TfeNet demonstrates optimal TD and BD in the selected two cases, successfully identifying almost all of the distal airways, and achieving a fine Precision. However, it also detected a higher number of incorrectly labeled airways at the airway termini, which contributes to its relatively lower Precision. UNet3D exhibited the fewest false positives (FP), achieving the highest Precision. However, its topological metrics are the worst, characterized by missed detections and fragmentation at the distal airway. WingsNet shows the most false negatives (FN) in the large airways, with noticeable airway erosion at the start of the large airways. Despite this, its effective loss function and network design enhances the detection of distal airways. Qin’s method exhibite a high number of false positives (FP) in both large and distal airways, particularly due to airway dilation, along with inadequate prediction capability for small airways.

\begin{figure}
    \centering
    \includegraphics[width=1\linewidth]{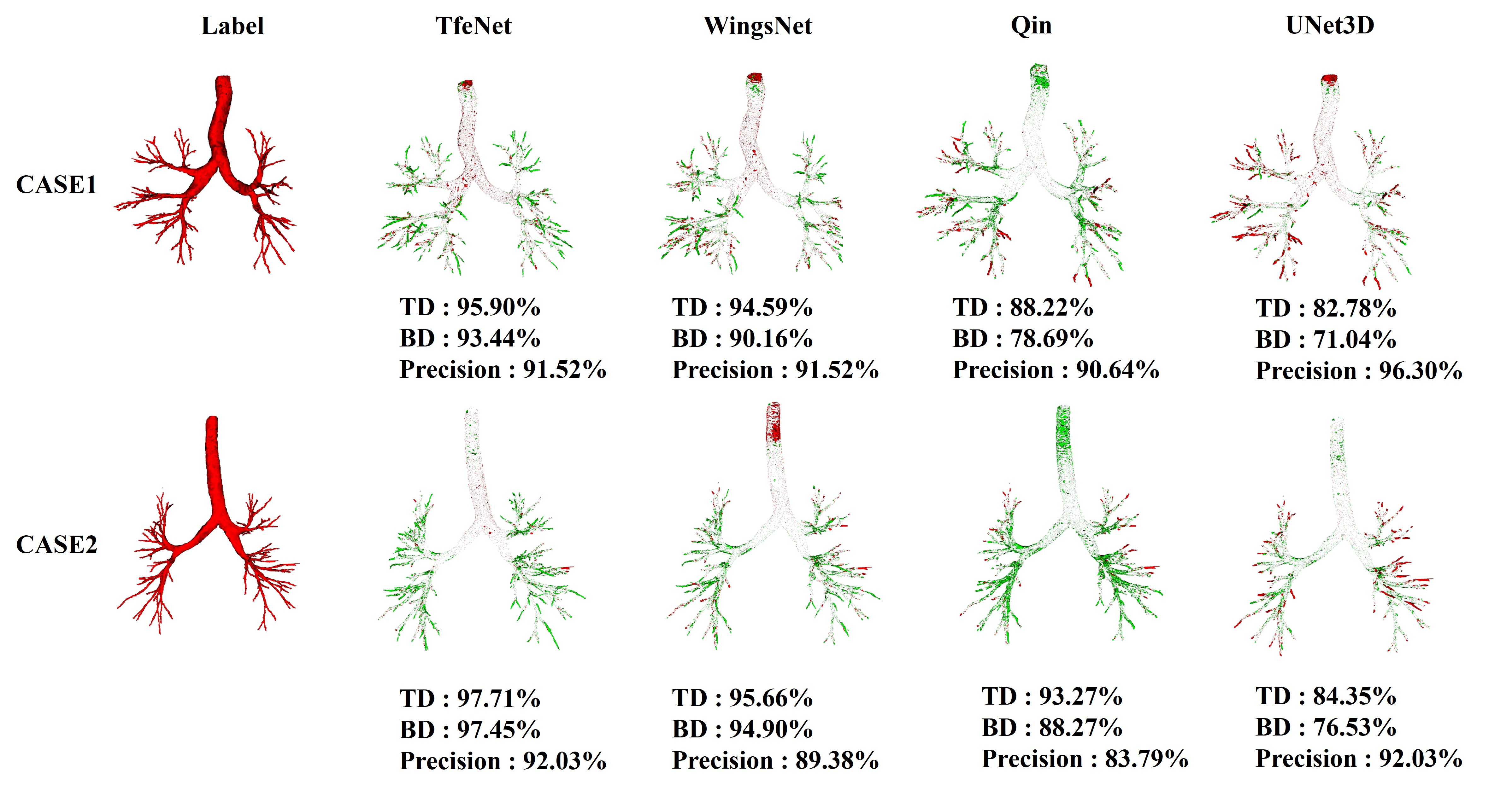}
    \caption{illustrates a visual comparison of advanced airway segmentation methods on the BAS test set. The first column displays the ground truth airway labels. The second column shows the segmentation results obtained by our proposed method. The remaining columns present the airway predictions from other methods, with false positives (FP) and false negatives (FN) highlighted in green and red, respectively.}
    \label{fig:enter-label}
\end{figure}

\subsection{Results on ATM22}
We submitted the predicted results of TfeNet on the validation set of the ATM22 challenge, as shown in Table 4. TfeNet achieved the highest DSC and MeanScore of 96.24\% and 96.21\%, respectively. LiuXinmeng achieved the highest Precision of 96.98\%, with TfeNet trailing by 1.25\% in Precision. However, TfeNet showed improvements in TD, BD, and DSC by 1.32\%, 2.21\%, and 0.34\%, respectively. Athon2 achieved the highest TD and BD scores of 97.61\% and 97.28\%, respectively, showing an improvement of 0.71\% and 1.03\% over TfeNet. However, their accuracy evaluation parameters, Precision and DSC, decreased by 3.42\% and 1.61\%, respectively, compared to TfeNet. In terms of enhancing branch length, they sacrificed a significant portion of accuracy. Overall, TfeNet demonstrated the most balanced topological and accuracy performance on the ATM22 validation set.

\label{}
\begin{table}
\caption{Quantitative comparison on ATM22 validation set. Besides TfeNet, all other baseline methods are taken from the Top-5 results on the validation set of ATM22 leaderboard. The best and next best results are in bold.}\label{tbl1}
\begin{tabular}{l c c c c c}
\toprule
Methods & TD(\%) & BD(\%) & Precision(\%) & DSC(\%) & MeanScore(\%) \\
\midrule
LiuXinmeng & $95.58 \pm 3.50$ & $94.13 \pm 5.30$ & $\mathbf{96.68 \pm 1.94}$ & $95.90 \pm 1.03$ & $95.57 \pm 2.94$ \\
nanda & $96.13 \pm 3.15$ & $94.29 \pm 4.89$ & $95.27 \pm 2.59$ & $\mathbf{96.01 \pm 1.19}$ & $95.43 \pm 2.96$ \\
haoyangCui & $96.43 \pm 3.24$ & $95.72 \pm 4.27$ & $94.55 \pm 2.62$ & $95.75 \pm 1.80$ & $\mathbf{95.61 \pm 2.98}$ \\
Cranberry & $96.29 \pm 3.01$ & $95.32 \pm 4.63$ & $95.00 \pm 2.77$ & $95.21 \pm 1.34$ & $95.46 \pm 2.94$ \\
athon2 & $\mathbf{97.61 \pm 2.58}$ & $\mathbf{97.28 \pm 3.10}$ & $92.01 \pm 2.92$ & $94.63 \pm 1.77$ & $95.38 \pm 2.59$ \\
\midrule
TfeNet(Ours) & $\mathbf{96.90 \pm 2.82}$ & $\mathbf{96.25 \pm 3.96}$ & $\mathbf{95.43 \pm 2.22}$ & $\mathbf{96.24 \pm 1.23}$ & $\mathbf{96.21 \pm 2.56}$ \\
\bottomrule
\end{tabular}
\end{table}

Additionally, we submitted our method in the form of a Docker container to the ATM22 organizing committee for offline testing. The test results of our proposed method and the advanced methods are shown in Table 5. Our method achieves the highest DSC and MeanScore of 95.49\% and 94.95\%, respectively. Moreover, our approach demonstrates the most balanced performance across all evaluated metrics. The difference between the highest metric (Precision) and the lowest metric (BD) in our method is only 2.2\%. Deeptree\_damo and Yby\_cas achieve the first and second highest TD and BD scores, with differences of 9.92\% and 4.65\% between their highest and lowest metrics, respectively. These results specifically reflect a strategy of sacrificing accuracy to gain higher topological performance. In contrast, NotBestMe and Iya achieve the first and second highest Precision scores of 96.59\% and 96.50\%, with differences of 15.25\% and 20.79\% between their highest and lowest metrics, respectively. This pattern indicates a trade-off in which topological integrity is compromised to enhance accuracy. Sanmed\_AI achieves the second highest DSC of 94.97\%, but its TD and BD performance is suboptimal. In summary, our method achieves the most balanced airway segmentation performance, maintaining competitive levels of both airway continuity and segmentation accuracy.

\label{}
\begin{table}
\caption{Quantitative comparison on ATM22 test set. Besides TfeNet, all other baseline methods are taken from the Top-10 results on the testing set of ATM22 leaderboard. The best and next best results are in bold.}\label{tbl1}
\begin{tabular}{l c c c c c}
\toprule
Methods & TD(\%) & BD(\%) & Precision(\%) & DSC(\%) & MeanScore(\%) \\
\midrule
Timi & $95.92 \pm 5.23$ & $94.73 \pm 6.39$ & $93.55 \pm 3.42$ & $93.91 \pm 3.68$ & $\mathbf{94.53 \pm 4.68}$ \\
Yby\_cas & $\mathbf{96.43 \pm 3.16}$ & $\mathbf{95.48 \pm 4.31}$ & $91.78 \pm 4.11$ & $93.83 \pm 1.90$ & $94.38 \pm 3.37$ \\
Yang & $94.51 \pm 8.60$ & $91.92 \pm 9.44$ & $94.71 \pm 8.30$ & $94.80 \pm 7.93$ & $93.99 \pm 8.57$ \\
DeepTree\_damo & $\mathbf{97.85 \pm 2.28}$ & $\mathbf{97.13 \pm 3.41}$ & $87.93 \pm 4.18$ & $92.82 \pm 2.19$ & $93.93 \pm 3.02$ \\
Neu204 & $90.97 \pm 10.41$ & $86.67 \pm 13.09$ & $93.03 \pm 8.41$ & $94.06 \pm 8.02$ & $91.18 \pm 9.98$ \\
Sanmed\_AI & $88.84 \pm 7.25$ & $83.35 \pm 10.90$ & $95.06 \pm 3.21$ & $\mathbf{94.97 \pm 1.80}$ & $90.55 \pm 5.79$ \\
Dolphins & $90.13 \pm 6.48$ & $84.20 \pm 11.15$ & $94.66 \pm 3.43$ & $92.73 \pm 2.09$ & $90.43 \pm 5.79$ \\
Suqi & $89.21 \pm 7.34$ & $82.16 \pm 12.26$ & $95.78 \pm 3.31$ & $93.65 \pm 2.10$ & $90.20 \pm 6.25$ \\
NotBestMe & $87.52 \pm 9.03$ & $81.34 \pm 13.56$ & $\mathbf{96.59 \pm 2.67}$ & $94.52 \pm 2.27$ & $89.99 \pm 6.88$ \\
Iva & $85.22 \pm 9.15$ & $75.71 \pm 14.89$ & $\mathbf{96.50 \pm 2.91}$ & $93.76 \pm 2.17$ & $87.80 \pm 7.28$ \\
\midrule
TfeNet(Ours)  & $95.34 \pm 4.17$ & $93.39 \pm 6.38$ & $95.59 \pm 3.26$ & $\mathbf{95.49 \pm 1.78}$ & $\mathbf{ 94.95 \pm 3.90}$ \\
\bottomrule
\end{tabular}
\end{table}

\subsection{Results on AIIB23}
To validate the performance of the proposed method on pulmonary fibrosis data, we conducted experiments on the AIIB23 dataset, as shown in Table 6. Since the standard deviation data for AIIB23 is not publicly available, only the average values of the evaluation parameters are presented in Table 6. On the AIIB23 validation set, our method achieves the second-highest Overall Score with a score of 89.26\%, demonstrating excellent airway segmentation performance. MinghuiZhang achieves the highest TD and BD scores of 90.54\% and 86.99\%, respectively, but their Precision, Leakage, and IOU parameters were lower than those of our method. Gexing achieved the highest IOU of 88.56\%, while TakashimaZakuro achieved the highest Precision and Leakage, with scores of 95.52\% and 96.06\%, respectively. However, their TD and BD metrics were suboptimal. Xiaoliu.ding and our method, though not achieving the best performance on individual metrics, ranked first and second in the Overall Score with 89.37\% and 89.26\%, respectively. In our Output1 stage, we used nnUNet for output, and when combined with Output2 from TfeNet, the TD and BD scores significantly increased by 10.45\% and 16.10\%, respectively. However, Precision, Leakage, and IOU slightly decreased by 3.81\%, 4.39\%, and 0.48\%, compared with nnUNet. The Overall Score increased by 2.58\%. In conclusion, our method achieved strong segmentation performance and led the field on fibrosis lung data, with significant improvements in airway continuity. It is important to note that in the Overall Score of the AIIB23 challenge, the Leakage parameter accounts for 30\%, emphasizing the importance of not overly focusing on TD and BD performance while neglecting FP.

\label{}
\begin{table}
\caption{Quantitative comparison on AIIB23 validation set. Besides TfeNet and nnUNet, all other baseline methods are taken from the Top-10 results on the validation set of AIIB23 leaderboard. The best and next best results are in bold.}\label{tbl1}
\begin{tabular}{l c c c c c c}
\toprule
Methods & TD(\%) & BD(\%) & Precision (\%) & IOU(\%) & Leakage(\%) & Overall Score(\%) \\
\midrule
Xiaoliu.ding & $\mathbf{90.50}$ & $\mathbf{86.56}$ & $91.04$ & $87.58$ & $90.41$ & $\mathbf{89.37}$ \\
twen & $88.28$ & $83.27$ & $92.61$ & $87.03$ & $92.42$ & $89.19$ \\
MinghuiZhang & $\mathbf{90.54}$ & $\mathbf{86.99}$ & $90.12$ & $86.43$ & $89.38$ & $88.78$ \\
Larry.Z & $87.72$ & $83.77$ & $91.44$ & $84.14$ & $91.36$ & $88.15$ \\
gexing & $81.29$ & $72.59$ & $94.46$ & $\mathbf{88.56}$ & $94.46$ & $87.29$ \\
TakashimaZakuro & $79.61$ & $70.16$ & $\mathbf{95.92}$ & $\mathbf{88.31}$ & $\mathbf{96.06}$ & $87.27$ \\
sanmed\_boyu & $83.20$ & $77.15$ & $92.42$ & $87.75$ & $92.14$ & $87.23$ \\
partho\_ghosh & $82.30$ & $74.48$ & $93.15$ & $85.52$ & $93.22$ & $86.67$ \\
Dolphins & $79.14$ & $69.92$ & $94.71$ & $87.32$ & $94.80$ & $86.38$ \\
segerard & $87.60$ & $82.47$ & $88.80$ & $83.69$ & $88.01$ & $86.35$ \\
nnUNet & $78.93$ & $69.56$ & $\mathbf{95.38}$ & $87.74$ & $\mathbf{95.51}$ & $86.68$ \\
\midrule
TfeNet(Ours) & $89.38$ & $85.66$ & $91.57$ & $87.26$ & $91.12$ & $\mathbf{89.26}$ \\
\bottomrule
\end{tabular}
\end{table}

\section{Discussion}
\subsection{Module ablation analysis}
This study sets up ablation experiments to verify the performance of the proposed module in airway segmentation. For fairness, a 7-layer 3D-UNet is used as the baseline, where the deconvolution module is replaced with a max-pooling module. Rotational data augmentation is applied with a threshold of 0.7, and the loss function is Tversky Loss with $\alpha$ set to 0.1. The kernel size of all linear convolutions and deformable convolutions is set to 7. The results of the ablation study are shown in Table 7, where "Linear" represents the use of linear convolution in the TFFM module, and DAConv45, DAConv60, DAConv90 represent the cases where q equals $\pi/4$, $\pi/3$, and $\pi/2$, respectively. When using linear convolution in the TFFM module, the accuracy metric slightly decreases compared to the baseline, but TD and BD increase by 1.58\% and 2.92\%, respectively, indicating the effectiveness of TFFM in improving airway continuity metrics. DSConv introduces continuity constraints into the transformation of convolution kernel sampling positions for the first time, enhancing the extraction ability for tubular targets. It also shows well performance in the TFFM module. Compared to Linear, TD and BD increase by 2.\% and 2.57\%, respectively. When q equals $\pi/2$, $\pi/3$, and $\pi/4$ in DAConv, TD, BD, and DSC gradually improve. The highest TD, BD, and DSC are achieved when $q=\pi/4$. Compared to DSConv, TD and BD increase by 0.97\% and 1.41\%, respectively, and compared to the baseline, both TD and BD improve by over 5\%. For a detailed analysis of $q$, please refer to Section 6.5 of the discussion.

\label{}
\begin{table}
\caption{Ablation study of the proposed method on the BAS test set. The best results are in bold.}\label{tbl1}
\begin{tabular}{l l c c c c}
\toprule
Methods & Convolution & TD(\%) & BD(\%) & Precision(\%) & DSC(\%) \\
\midrule
Baseline w/o TFFM & $\times$ & $89.77 \pm 10.43$ & $85.96 \pm 12.06$ & $\mathbf{81.02 \pm 4.64}$ & $88.10 
\pm 2.29$ \\
\midrule
 & Linear & $91.35 \pm 13.08$ & $88.88 \pm 14.53$ & $80.69 \pm 4.93$ & $87.42 \pm 3.77$ \\
 & DSConv \citep{Qi2023DynamicSC} & $93.67 \pm 5.47$ & $91.45 \pm 7.23$ & $80.42 \pm 4.82$ & $87.78 \pm 2.82$ \\
TfeNet w TFFM & DAConv45 & $\mathbf{94.64 \pm 6.21}$ & $\mathbf{92.86 \pm 7.81}$ & $80.49 \pm 4.95$ & $\mathbf{88.13 \pm 2.58}$ \\
& DAConv60 & $93.09 \pm 10.41$ & $90.82 \pm 11.96$ & $80.53 \pm 4.96$ & $87.71 \pm 3.24$ \\
& DAConv90 & $91.83 \pm 11.51$ & $89.59 \pm 13.11$ & $80.55 \pm 4.91$ & $87.41 \pm 3.67$ \\
\bottomrule
\end{tabular}
\end{table} 

\subsection{False positives}
On the BAS dataset, the proposed TfeNet demonstrates competitive topological metrics, though its precision is approximately 7\% lower than V-Net. As visualized in Figure 5(a-d), false positive (FP) errors predominantly cluster in small airways and manifest in two patterns: Firstly, incomplete reference labels (Figure 5c-d), where region-growing based annotations lack refinement of distal branches; Secondly,  conservative boundary labeling to avoid lumen over-expansion (Figure 5a-b). Notably, such mild lumen dilation has minimal impact on clinical utility, making these FPs clinically acceptable. This precision discrepancy is substantially mitigated on the ATM22 dataset with higher annotation quality, where TfeNet achieves leading performance in both continuity and accuracy metrics.

\begin{figure}
    \centering
    \includegraphics[width=0.5\linewidth]{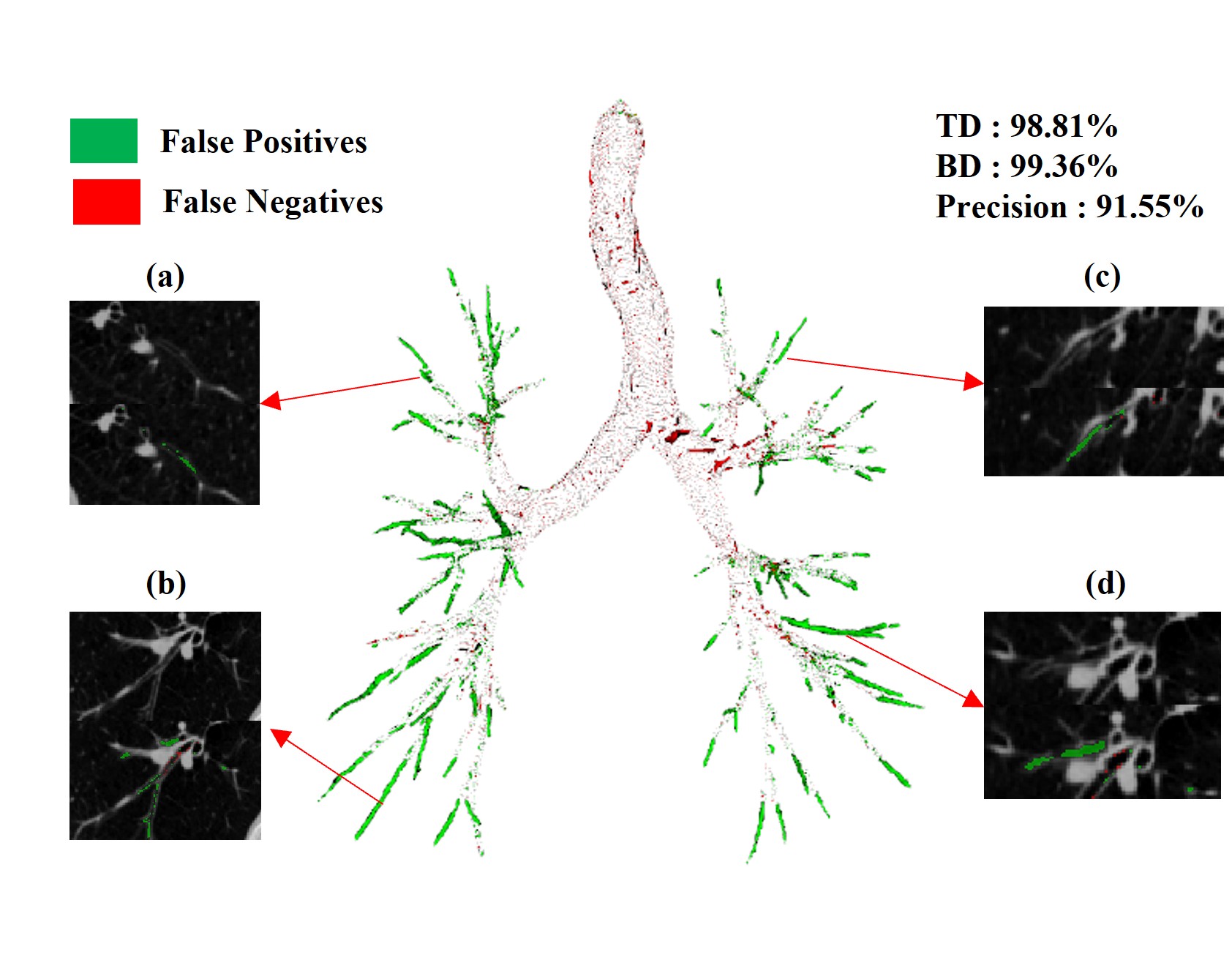}
    \caption{Analysis of false positive results in airway segmentation on the BAS dataset. Four key regions of false positives were selected. The first row shows the original CT images, and the second row shows the analyzed images. In both the analyzed images and 3D views, false positives are marked in green, while false negatives are marked in red.}
    \label{fig:enter-label}
\end{figure}

\subsection{False negatives}
We selected a case from the BAS test set with the most significant false negatives (FN) for analysis. The most notable impact of increased FN is the severe disruption of airway continuity, with a TD and BD of only 82.53\% and 78.44\%, respectively, in the selected example. Figure 6 shows the false negative labels for several key airway locations: (a), (b), (c), and (d). FN can be categorized into three types. The first type of FN is the missed detection of entire small airways, as shown in Figure 5c-d. The primary cause of this type is the CT scan resolution, which causes the CT values of the airways to be similar to those of surrounding tissues. The second type is due to the blurring around the walls of large airways, leading to slight constriction, as shown in Figure 6 (c). The third type is caused by uneven CT intensity distribution within the airway, resulting in small airway discontinuities, as seen in Figure 6 (b). After extracting the largest connected region, these areas will be discarded. In general, the main factors contributing to these FNs are, on the one hand, the resolution limitations of CT equipment, and, on the other hand, the small training set of the BAS dataset (50 cases), which is insufficient to generalize airway characteristics under different conditions.

\begin{figure}
    \centering
    \includegraphics[width=0.5\linewidth]{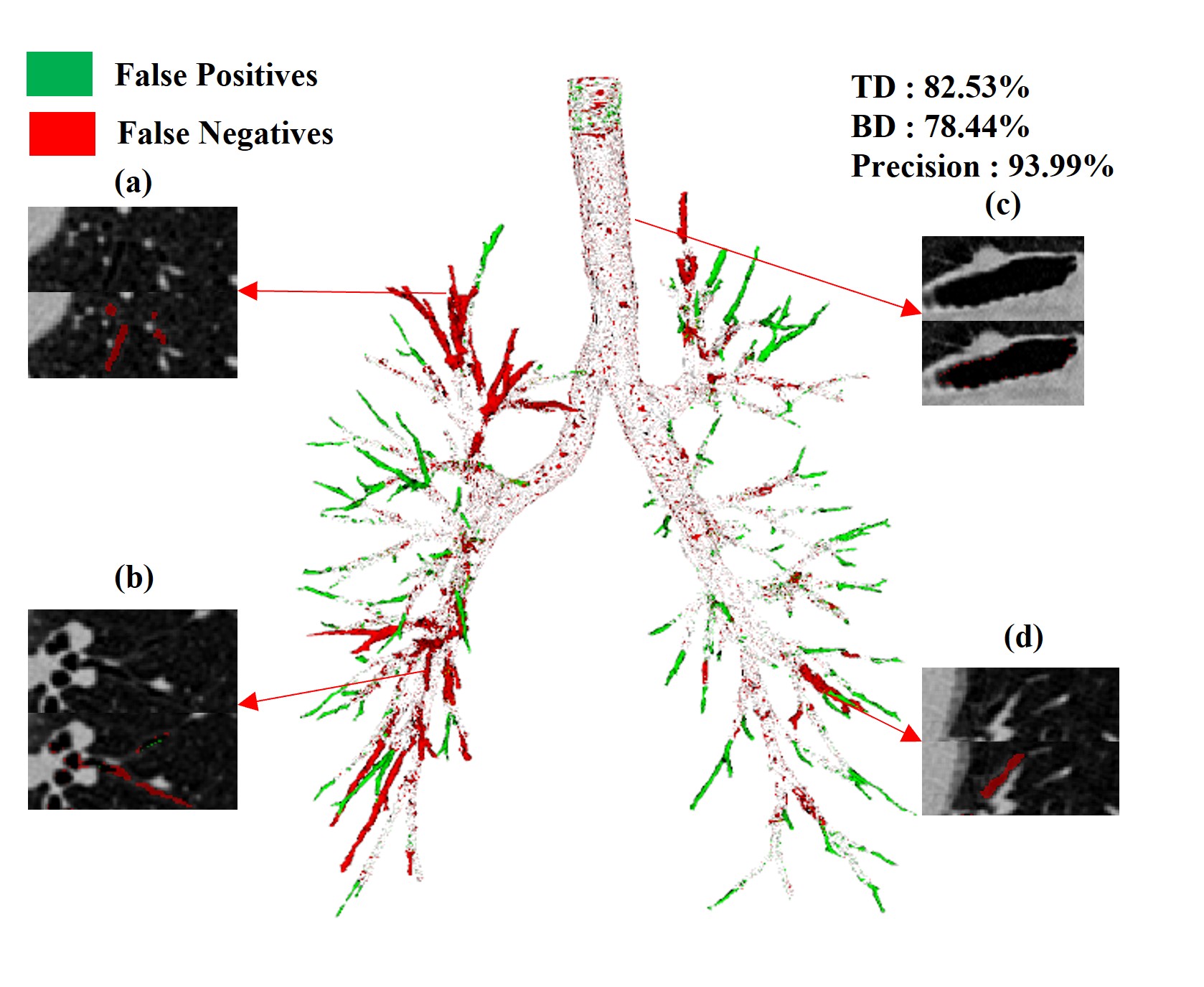}
    \caption{Analysis of false negatives results in airway segmentation on the BAS dataset. Four key regions of false negatives were selected. The first row shows the original CT images, and the second row shows the analyzed images. In both the analyzed images and 3D views, false positives are marked in green, while false negatives are marked in red.}
    \label{fig:enter-label}
\end{figure}

\subsection{Features enhancement of DAConv}
The DAConv we propose enhances the focus on fine airway features. In Figure 7, each row represents the convolution shapes and convolution heatmaps along the x, y, and z directions. From the x-direction slice, the center of the convolution kernel is positioned within the dot-like airway structure, with the kernel uniformly distributed around the structure. The maximum value of the heatmap coincides with the center of the airway. Similarly, from the slices in the y and z directions, the sampling positions of the convolution kernel are evenly distributed along the direction of the airway extension, and the maximum values of the heat maps are aligned with the linear airway structures. In general, our method improves the features of the airway structure in different directions of slice, helping the network capture more useful features.

\begin{figure}
    \centering
    \includegraphics[width=0.75\linewidth]{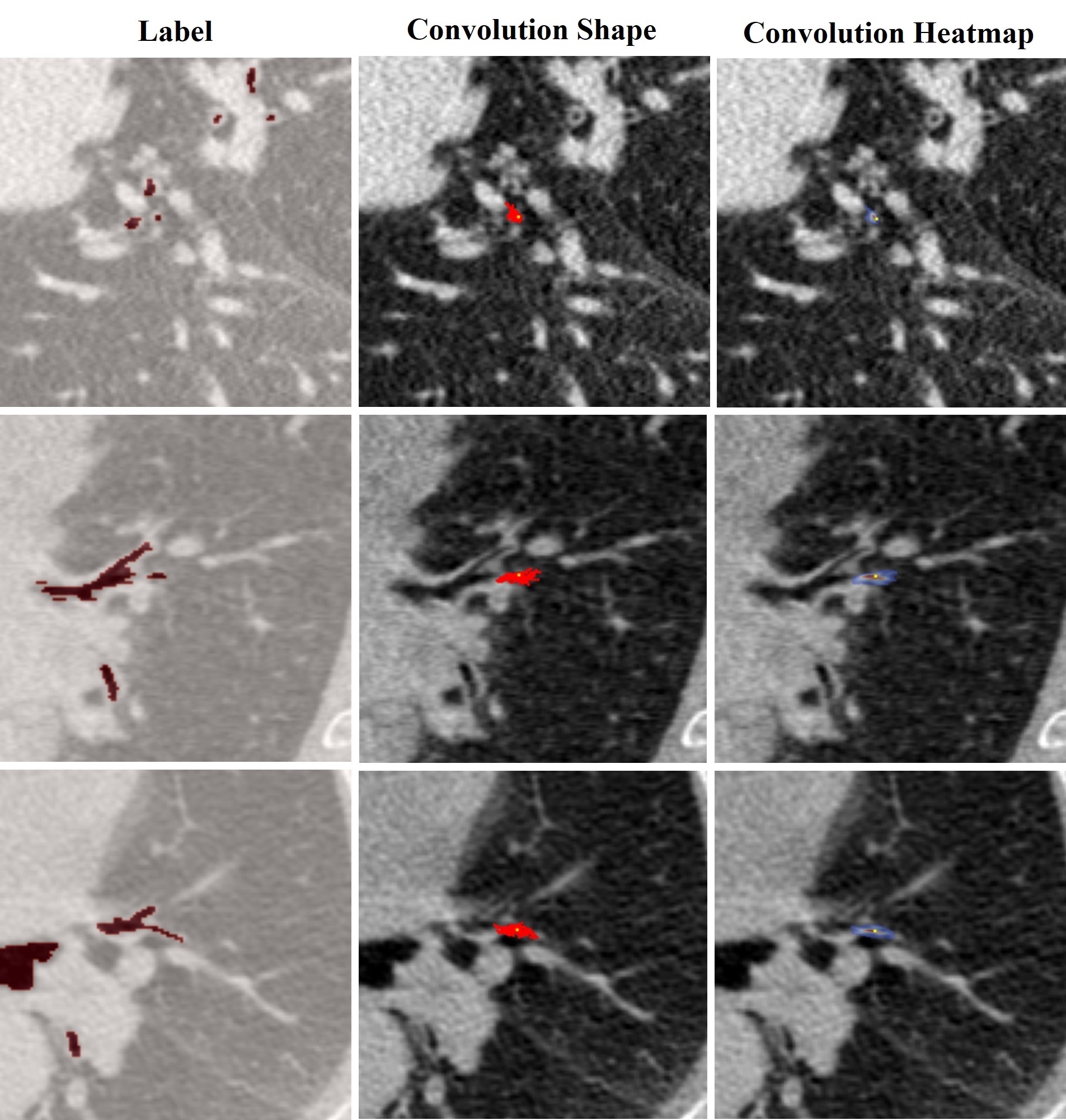}
    \caption{DAConv Sampling Locations. The yellow dot represents the center of the selected convolution kernel, while the red dots indicate the sampling positions of the 57 (1+7+49) convolution kernels related to the center kernel. The heatmap shows the overlap degree of these convolution kernel sampling positions.}
    \label{fig:enter-label}
\end{figure}

\subsection{Perspectives on deformable linear convolutions}
In this study, we select $q=\pi/4$, and the DAConv sampling spaces in different directions are illustrated in Figure 3(c). The low overlap at the edges helps to avoid learning unnecessary or redundant features. In three-dimensional medical image segmentation tasks, considerations for deformable convolutions akin to DSConv and DAConv suggest that not all orientations of deformable linear convolutions necessarily benefit the network performance. Notably, our experiments reveal that removing residual connections from the TFFM led to an over-representation of redundant local features, deteriorating the training effectiveness of the network, and a similar issue is observed with DSCNet. Furthermore, airway trees exhibit distinct direction-aware characteristics; the growth directions differ significantly between the upper left lobe and the lower left lobe, as well as between the upper right lobe and the lower right lobe. A key direction for future research is how to appropriately weight deformable linear convolutions in different directions. Implementing a reasonable dropout training strategy or employing angle parameter learning with adequate supervision signals could help suppress redundant features while enhancing the strength of useful features. These strategies aim to refine feature extraction by focusing on meaningful structural information and improving the overall segmentation accuracy for complex tubular networks such as airways. This approach will be critical in the advancement of the field of medical image segmentation toward more precise and reliable outcomes.

\section{Conclusion}
In this work, a novel airway segmentation network named TfeNet is proposed. TfeNet introduces a novel direction-aware convolution operator, which achieves spatial deformation of linear convolution through a learnable angle layer, thereby enhancing the strength of airway region features. Based on asymmetric convolution and residual connections, a tubular feature fusion module is designed to preserve input information while integrating local tubular features, ensuring information flow and expanding the receptive field. Extensive experiments are conducted on a public airway dataset and two challenges. The experimental results demonstrate that the proposed method achieves significant advantages in balancing airway topology improvement and precision. The results also highlight the great potential for application of our airway segmentation approach in bronchoscopic navigation and robotics.

\printcredits

\section*{CRediT authorship contribution statement}

Qibiao WU: Methodology, Writing - Original Draft. Yagang WANG: Project administration, Writing - Review, , Funding acquisition \& Editing. Qiang ZHANG: Writing - Review and Editing, Funding acquisition.

\section*{Declaration of Competing Interest}

The authors declare that they have no known competing financial interests or personal relationships that could have appeared to influence the work reported in this paper.

\section*{Acknowledgment}

This research is partly supported by the National Natural Science Foundation of China: 62206114. 

\bibliographystyle{cas-model2-names}

\bibliography{cas-refs}





\end{document}